
\input harvmac
\Title{\vbox{\baselineskip12pt\hbox{IC/93/236}}}
{\vbox{\centerline{Gauging of Chern-Simons $p$-Branes}}}


\centerline{Raiko P. Zaikov{\footnote{$^\dagger$}{Permanent address:
Institute for Nuclear Research and Nuclear Energy, Boul.
Tzarigradsko Chaussee 72, 1784 Sofia, Bulgaria, e-mail:
zaikov@bgearn.bitnet}}
\footnote{$^\star$}
{Supported by Bulgarian Fondation on Fundamental Research under
contract Ph-11/91-94}}
\bigskip\centerline{International Centre for Theoretical Physics}
\centerline{Strada Costeiera 11, P. O. Box 11, Trieste, Italy}
\centerline{}

\vskip .3in

\noindent{\bf Abstract.} The Chern-Simons membranes and in
general the Chern-Simons p-branes moving in $D$-dimensional
target space admit an infinite set of secondary constraints. With
respect to the Poisson bracket these constraints form a closed
algebra which contains classical \ $W_{1+\infty }$ \ algebra in
$p$-dimensions as a subalgebra. Corresponding gauged theory in
the phase-space is constructed in a Hamilton gauge as an analog
of the ordinary $W$-gravity.

\Date{7/93} 


\noindent In the previous article \ref\rZ{R. P. Zaikov, {\it
"Chern-Simons p-branes and p-dimensional W-algebras"}, INRNE
preprint INRNE-TH/3/93, hep-th/9304075} it was shown that in the
Chern-Simons membrane case there always appears an infinite set
of secondary constraints in contrast to the C-S string theory
\ref\rSZa{M. N. Stoilov and R. P. Zaikov, Lett.  Math. Phys., {\bf
27}, 155 (1993).} in which there are two possibilities for the
first class constraints : there is a finite or an infinite number
of secondary constraints. There is also another, rather formal,
possibility when second class constraints appear also (see Refs.
1  and 2
\nref\rSZ{M. N. Stoilov and R. P. Zaikov, Mod. Phys. Lett., {\bf
8}, 313 (1993).}).

When there appears an infinite set of secondary constrains
for the C-S string, they satisfy an infinite algebra
with respect to the Poisson bracket. This algebra contains as a
subalgebra the classical (without central term) affine $SL(2,R)$
algebra, as well as the classical Virasoro algebra and
their higher spin extensions which contain classical $W_{1+\infty }$
algebra \ref\rP{C. N. Pope, L. J. Romans and X. Shen, Pys. Lett.
{\bf B 336}, 173 (1990) and {\bf 242B}, 401 (1990).}. Through this
paper we are dealing only with classical infinite algebras.

In the case of C-S membrane the infinite set of constraints gives
a linear realization of higher spin extended algebra in
two-dimensions which contains affine $SL(2,R)$, Virasoro and
$W_{1+\infty }$-algebras in two dimensions as subalgebras\foot{In
ordinary two-dimensional conformal theories we have two copies of
Kac-Moody, Virasoro and $W$ algebras each of which acts only on
one light-cone (holomorphic or antiholomorphic) coordinate, i.e.
we have direct product of two algebras in one-dimension.}\rZ .
We note, that any of these algebras can not be represented as a
direct product of two infinite algebras in one dimension as in
the two-dimensional conformal theory
\ref\rFR{J. M. Figueroa-O'Farrill and E. Ramos, Phys. Let., {\bf
B 282}, 357 (1992).}
\nref\rMR{F. Martinez-Moras and E. Ramos, {\it Higher Dimensional
Classical $W$-algebras}, Preprint-KUL-TF-92/19 and US-FT/6-92,
(hep-th/9206040)}--\ref\rMMR{F. M. Moras, J. Mas and E. Ramos,
{\it Diffeomorphisms from higher dimensional $W$-algebras},
Preprint, QMW-PH-93-1, US-FT/1-91, (hep-th/9303034)} and their
$W_\infty $ \ extensions\foot{In the present article we use the
terminology of the papers \rMR \ and
\rMMR .}. We have to take into account also that in the C-S membrane
theory we are dealing with two spatial dimensions while the time
variable appears only as evolution parameter which is not the
case on the ordinary 2D conformal theory. The generalization of
the results for arbitrary C-S $p$-branes is strigtforward. In
that case we have higher spin extension of the affine  $SL(2,C)$
in $p$-dimensions.

The polynomial Chern-Simons $p$-brane action was obtained
in \ref\rZa{R. P. Zaikov, Phys. Lett., {\bf B 266}, 303 (1991).} from
the topological $(p+1)$-brane action (only for \ $D=p+2$)
\ref\rF{B. Biran, E. G. Floratos and G. K. Savvidi, Phys. Lett.
{\bf B 198}, 329 (1987).}\nref\rZz{R. P. Zaikov, Phys. Lett. {\bf B
211}, 281 (1988).}\nref\rGTz{M. P. Grabowski and C-H. Tze, Phys.
Lett. {\bf B 224}, 259 (1989).} --\ref\rFKT{K. Fudjikawa, J. Kubo
and H. Terao, Phys. Lett., {\bf B 263}, 371 (1991).} in the same
way as in the ordinary local theory.  In the paper \ref\rZc {R.
P. Zaikov, Phys.  Lett., {\bf B 263}, 209 (1991).} a generalization
for an arbitrary space-time dimension was found.

In the present article the gauged C-S mebrane theory is
constructed in the Hamiltonian approach considering the Lagrange
multiplyers as a gauge fields with arbitrary spin. The
$W_{1+\infty }$ transformation properties of the gauge fields are
obtained. We note that the {\it "No go theorem"} in the case of
spin $>2$ (see Ref. \ref\rBB{F. A. Berends and G. J. H. Burgers,
Nucl.  Phys. {\bf B260}, 295 (1985).}) does not take place because
we are dealing with an infinite sequence of higher spin gauge
fields.  Through this paper we use basis in which not all the
constrains are independent and as a consequence there exists an
additional symmetry of Stuckelberg type \ref\rBPSS{E. Bergshoeff,
C. N. Pope, L. J. Romans, E. Sezgin, X. Shen and K. S. Stelle,
Phys. Lett., {\bf B 243}, 350 (1990).}. This symmetry allows us
to exclude the corresponding gauge fields (with odd spin) by
means of gauge fixing procedure. Integrating over the momentum
variables we find $W_{1+\infty }$ gauge invariant action on the
configuration space. We remaind that the Lagrangian approach to
the $W$-gravity was considered in a lot of papers among the first
of which are \ref\rH{C. M. Hull, Phys. Lett., {\bf B 240}, 110 (1990);
 Nucl. Phys., {\bf B364}, 62 (1991).} and \rBPSS .

To proceed further we shall remind brifly some results from the
papers \rZa \ and \rZc , where in order to generalize the C-S
$p$-brane to arbitrary space-time dimension the following
notation were introduced:
$$X_{,A}=\partial _AX=\matrix{ X
&  {\rm if}  & A=* \cr X_{,a}=\partial _aX &  {\rm if}  & A=a,
\cr }$$
($a=0,1,2,\dots ,p$). The polynomial Lagrangian for the
C-S $p$-brane is given by  $L=detX^\mu _{,A}$ which exists in  \
$D=p+2$ \ only. To extend this action for target space with
arbitrary dimension,  a generalized induced metric tensor is
introduced
\eqn\ew{\widetilde g_{AB}=X^\mu _{,A}X^\nu _{,B}\eta _{\mu \nu },}
where $\eta _{\mu \nu }$ is the pseudoeuclidean metric tensor.
{}Formally replaciment of the ordinary induced
metric tensor in the Nambu-Goto action with the generalized
metric tensor given by Eq. \ew \ gives the
action for the generalized C-S $p$-branes wich lives on a target
space with arbitrary dimension
\eqn\eal{S=
\kappa \int d\tau d^p\sigma
\sqrt{-\widetilde g}.}

It is easy to check that the action \eal \ obei the same
$p+1$-variable diffeomorphisms invariance as the ordinary
p-branes action. As a consequence of this invariance from the
following first class primary constraints are obtained:
\eqn\eca{\eqalign{
\phi _\perp &=
{\cal P}^2+\kappa ^2det\Bigl(X_{,u}X_{,v}\Bigr)\approx 0, \cr
\phi _j &={\cal P}X_{\sigma _j}\approx 0, \cr
\phi _* &={\cal P}X\approx 0,\cr }}
where $u,v=*,1,\dots ,p; \ j=1, \dots ,p$. Hence, there appears
one additional primary constraint and moreover, the degree of the
first constraint is higher by two degrees than the degree of the
corresponding constraint in the ordinary $p$-branes.  The latter
is a consequence of the fact that the Lagrangian given by Eq.
\eal \ and the constraints \eca
\ are obtained from the corresponding ordinary $(p+1)$-brane
Lagrangian  and $(p+1)$-brane constraints  substituting $\partial
_{\sigma _{p+1}}X$ by $X$. We recall, that the ordinary bosonic
string has only two (bilinear) constraints while the C-S string
has three primary costraints, one of which is of degree four with
respect to $X$.

The appearance of the constraint $\phi _*$ shows us that some
residual symmetry from the $p+2$-variable diffeomorfisms (under
which the action of the $p+1$-brane is invariant) survives.  As a
consequence of the appearance of the constraint $\phi _*$ there
arise some secondary
constraints too. In the C-S particle case we have only one secondary
constraint, while in the C-S string case  there are two
possibilities: four (three primary
constraints and one secondary constraint) first class constraints
\rSZ \ or an infinite set of first
class constraints \rSZa . We note that in the latter case not all
of the constraints are independent if we deal with finite
dimensional target space. Hence we have not dynamical degree of
freedom. In that case the
dynamical degrees of freedom can take place only if we have
infinite dimensional target space.

For any  C-S $p$-brane the canonical Hamiltonian vanishes
identically, i.e.
\eqn\ewd{{\cal H}_0={\cal P}\dot X -{\cal L}\equiv 0,}
which is a property of the ordinary p-brane theory also.

The analyze of the constraint algebra in the case of C-S
membrane shows  us that there is an infinite
series of secondary constraints \rZ . An appropriate choice of
these constraints is the following:
\eqn\ech{\eqalign{\Psi ^{m,n} & =\Bigl({\cal P}\partial _{\sigma _1}^m
\partial _{\sigma _2}^n{\cal P}\Bigr)\approx 0, \cr
\Phi ^{m,n} & =\Bigl(X\partial _{\sigma _1}^m\partial _{\sigma
_2}^nX\Bigr)\approx 0, \cr
\Gamma ^{m,n} & =\Bigl({\cal P}\partial _{\sigma 1}^m\partial _{\sigma
_2}^nX\Big)\approx 0, \qquad  (m,n=0,1,\dots ). \cr }}

We note that, as it was mentioned above, if $D$ is finite
we have only a finite number of independent constraints \ech .
 However,  when we are dealing with infinite dimensional target
space it is easy to check that all the
constraints \ $\Gamma $ \ are independent as well as those of the
constraints $\Psi $ and $\Phi $ for which $m+n=2k$ ($k=0,1,\dots
$). To prove the latter statement we use the following identity
\eqn\ewa{(\partial ^m_{\sigma _1}\partial ^n_{\sigma _2}XY)=
\sum _{p=0}^m\sum _{q=0}^n(-)^{m+n-p-q}\pmatrix{m \cr p \cr }
\pmatrix{n \cr q \cr }\partial ^p_{\sigma _1}\partial ^q_{\sigma _2}
(X\partial ^m_{\sigma _1}\partial ^n_{\sigma _2}Y)}
which consequend from the Laibniz formula.

Using the  Eq. \ewa \ we obtain that the constraints
$\Psi $ and \ $\Phi $ \ with arbitrary odd spin can be
represented in terms of the constrains with all underlying
spins:
\eqn\ewb{(XX^{2k-l+1,l})=
{1\over 2}\sum _{p=0}^{2k-l+1}\sum
^l_{q=0, q+p\ne 0}(-)^{p+q+1}\pmatrix{2k-l+1 \cr p \cr }
\pmatrix{l \cr q \cr }\partial^p_{\sigma _1}\partial ^q_{\sigma _2}
(XX^{2k-l-p+1,l-q}),}
where $k,l=0,1,2,\dots $. To obtain the r.h.s. of the Eq.
\ewb \ for given $k,l$ only in terms of
independent quantities we have to determine all of the even lower spin
quatities from the corresponding equation and then to insert
them in
the r.h.s. of \ewb . In such a way we get
\eqn\edk{\bigl(XX^{2K-l+1,l}\bigr)=\sum _{M,m}C_{M,m}^{K,l}
\partial _{\sigma _1}^{2(K-M)-l+m+1}\partial _{\sigma _2}^{l-m}
\bigl(XX^{2M-m,m}\bigr),}
where the sumation over \ $m$ \ is from \ $0$ \ to \
 $min(l, 2M)$ \ and over \ $M$ \ is from \ $0$ \ to \
 $K+(m-l+1)/2$. The coefficient
\ $C^{K,l}_{M,m}$ \ can be determined by the procedure described
above.

Althout, only the constraints \ $\Psi $ \ and \ $\Phi $ \ for which
\ $m+n=2k$ \ are independent for convenience we do not exclude the
odd spin constraints, moreover, that all the constraints \
$\Gamma $ \ are independent (for $D=\infty $).

With respect to the Poisson bracket the constraints \ech \ form
an infinite algebra which contains \ $W_{1+\infty }$
\ algebra in two-dimensions as a subalgebra \rZ :
\eqn\ecl{\eqalign{
\{\Gamma ^{k,l}\bigl[f\bigr],\Gamma ^{m,n}\bigl[h\bigr]\}_{PB}=
 & \sum _{p=0}^k\sum _{q=0}^l
\pmatrix{k \cr p \cr}\pmatrix{l \cr q \cr }
\Gamma ^{k+m-p,l+n-q}\bigl[f
\partial _{\sigma _1}^{p}\partial _{\sigma _2}^{q}h\bigr] \cr
& -\sum _{r=0}^m\sum _{s=0}^n
\pmatrix{m \cr r \cr}\pmatrix{n \cr s \cr }
\Gamma ^{k+m-r,l+n-s}\bigl[
h\partial _{\sigma _1}^r\partial _{\sigma _2}^sf\bigr]. \cr }}

We note, that Eq. \ecl \ contains as subalgebras also two copies
of $W_{1+\infty}$ algebras in linear realization -- one of which
acts on $\sigma _1$ coordinate and the other one acts on $\sigma
_2$ coordinate\foot{These $W_{1+\infty }$-algebras differ from
the ordinary $W_{1+\infty }$-algebras because they are not
mutually commuting.}. The constraints $\Gamma ^{k,0} \ {\rm and}
\ \Gamma ^{0,k}$ appear as generators of these transformations.
In the general case $\Gamma ^{k,l}$ can be considered as
generators of generalized diffeomorphismes in two dimensional
space. Indeed, the Poisson bracket of $\Gamma $ with the
coordinate $X^\mu $ \ and with the momenta \ ${\cal P}^\mu $ give
the transformation laws for the phase space coordinates:
\eqn\ecn{\eqalign{\delta _\Gamma ^{k,l}X^\mu =
\{\Gamma ^{k,l}\bigl[f\bigr],X^\mu \}_{PB} & =
-f\partial _{\sigma _1}^k\partial _{\sigma _2}^lX^\mu ,\cr
\delta _\Gamma ^{k,l}{\cal P}^\mu =
\{\Gamma ^{k,l}\bigl[f\bigr],{\cal P}^\mu \}_{PB} & =
(-)^{k+l}\sum _{p=0}^k\sum _{q=0}^l\pmatrix {k \cr p \cr }
\pmatrix {l \cr q \cr }\partial _{\sigma _1}^p\partial _{\sigma
_2}^qf\partial _{\sigma _1}^{k-p}\partial _{\sigma_2}^{l-q}
{\cal P}^\mu .\cr }}
In the same way we obtain also:
\eqn\eco{\eqalign{ & \delta _{\Phi }^{k,l}X^{\mu }=
\{\Phi ^{k,l}\bigl[f\bigr],X^{\mu }\}_{PB}=0, \cr
&\delta _{\Phi }^{k,l}{\cal P}^{\mu }=
\{\Phi ^{k,l}\bigl[f\bigr],{\cal P}^{\mu }\}_{PB} \cr
& =-f\partial _{\sigma _1}^{k}
\partial _{\sigma _2}^{l}X^{\mu }
-(-)^{k+l}\sum _{p=0}^{k}\sum _{q=0}^{l}
\pmatrix{k \cr p \cr }\pmatrix{l \cr q \cr }
\partial _{\sigma _1}^{p}
\partial _{\sigma _2}^{q}f\partial _{\sigma _1}^{k-p}
\partial _{\sigma _2}^{l-q}X^{\mu }, \cr
& \delta _{\Psi }^{k,l}X^{\mu }=
\{\Psi ^{k,l}\bigl[f\bigr],X^{\mu }\}_{PB} \cr
& =f\partial _{\sigma _1}^{k}
\partial _{\sigma _2}^l{\cal P}^{\mu }(-)^{k+l}
\sum _{p=0}^k\sum _{q=0}^l
\pmatrix{k \cr p \cr }\pmatrix{l \cr q \cr }
\partial _{\sigma _1}^{p}\partial _{\sigma _2}^{q}f
\partial _{\sigma _1}^{k-p}
\partial _{\sigma _2}^{l-q}{\cal P}^{\mu }, \cr
& \delta _{\Psi }^{k,l}{\cal P}^{\mu }=
\{\Phi ^{k,l}\bigl[f\bigr],{\cal P}^{\mu }\}_{PB}=0. \cr }}
Consequently, $\delta _\Gamma ^{1,0} \ {\rm and} \ \delta _\Gamma
^{0,1}$ are ordinary diffeomorphisms in two-dimensional
space.  We note, that the assymmetry which appears in the
transformation laws of the coordinate $X$ and momentum ${\cal P}$
is a consequence of the assymetric choise of the constraint
basis \ech .  Taking into account the identity \ewa \
a more symmetric basis can be obtained for the constraints
\ech \ by a simple redefinition
$$
\Lambda ^{m,n}\rightarrow \widetilde \Lambda ^{m,n}=
\sum _{p=0}^m\sum _{q=0}^{n}b^{mn}_{pq}\partial _{\sigma _1}^p
\partial _{\sigma _2}^{q}\Lambda ^{m-p,n-q},
$$
where $b$ are constants. By a suitable choise
of  $b$ the classical algebra \ecl \ can be
deformed to an algebra which admits diagonal central extension
 \rP \ also,  at least for the $W_{1+\infty }$ subalgebra.

Because of the vanishing of the canonical Hamiltonian given by
Eq. \ewd \ the first order action can be writen in the form:
\eqn\eda{S=\int d\tau d^2\sigma \biggl({\cal P}{\dot X}-
\alpha _{mn}\bigl(XX^{(m,n)}\bigr)-
\beta _{mn}\bigl({\cal P}{\cal P}^{(m,n)}\bigr)-
\gamma _{mn}\bigl({\cal P}X^{(m,n)}\bigr)\biggr),}
where  $U^{(m,n)}=\partial ^m_{\sigma _1}
\partial ^n_{\sigma _2}U$.
In order to gauge the action givev by Eq. \eda \ we consider the
lagrange multiplyers \ $\alpha ,\beta \ {\rm and} \ \gamma $ \ as
fields depending on the evolition parameter $\tau $ also. Then
using the transformation laws for the phase-space coordinates
given by Eqs. \ecn \ and \eco \ with $\tau $ depending parameters
$f$ we obtain the transformation laws for the gauge fields:
\eqn\edb{\eqalign{\delta _{\Gamma }\alpha _{mn} & =
\sum _{k,l\ge 0}\sum _{r,s=0}^{k,l}(-)^{r+s}\pmatrix{k \cr
r \cr }\pmatrix{l \cr s \cr }\partial _{\sigma _1}^r
\partial _{\sigma _2}^s\biggl(f_{kl}\alpha _{m-k+r,n-l+s}\biggr) \cr
& +\sum _{k,l\ge 0}\sum _{r,s=0}^{k,l}\pmatrix{k \cr
r \cr }\pmatrix{l \cr s \cr }\alpha _{kl}\partial _{\sigma _1}^r
\partial _{\sigma _2}^sf_{m-k+r,n-l+s}, \cr }}
\eqn\edc{\eqalign{\delta _{\Gamma }\beta _{mn} & =
-\sum _{k,l\ge 0}\sum _{r,s=0}^{k,l}(-)^{k+l}\pmatrix{k+r \cr
r \cr }\pmatrix{l+s \cr s \cr }f_{kl}\partial _{\sigma _1}^r
\partial _{\sigma _2}^s\beta _{m-k+r,n-l+s} \cr
& -\sum _{k,l\ge 0}\sum _{r,s=0}^{k,l}(-)^{k+l}\pmatrix{k \cr
r \cr }\pmatrix{l \cr s \cr }\beta _{kl}\partial _{\sigma _1}^r
\partial _{\sigma _2}^sf_{m-k+r,n-l+s}, \cr }}
\eqn\edd{\eqalign{\delta _{\Gamma }\gamma _{mn} & =
-{\dot f}_{mn}-\sum _{k,l\ge 0}\sum _{r,s=0}^{k,l}\pmatrix{k \cr
r \cr }\pmatrix{l \cr s \cr }f_{kl}\partial _{\sigma _1}^r
\partial _{\sigma _2}^s\gamma _{m-k+r,n-l+s} \cr
& -\sum _{k,l\ge 0}\sum _{r,s=0}^{k,l}\pmatrix{k \cr
r \cr }\pmatrix{l \cr s \cr }\gamma _{kl}\partial _{\sigma _1}^r
\partial _{\sigma _2}^sf_{m-k+r,n-l+s}, \cr }}

We note, that the action \eda \ is invariant only with respect to
the gauged $W_{1+\infty }$ algebra in two dimensions.  It is not
invariant with respect to the local gauge transformations \eco .

In order to write down the action \eda \ on the configuration
space we exclude the momentum variables by means of the equation:
\eqn\ede{{\delta{\cal L}\over \delta {\cal P}}=0.}
Incerting the Lagrangian from \eda \ into Eq. \ede \ we find
\eqn\edg{\dot X_\mu-\alpha X_\mu ^{(m,n)}-{\cal Q}{\cal P}_\mu =0,}
where
\eqn\edf{{\cal Q}={1\over 2}\sum _{m,n\ge 0}\biggl(\beta _{mn}
\partial _{\sigma _1}^m\partial _{\sigma _2}^n+
\sum _{p,q=0}^{m,n}\partial _{\sigma _1}^p\partial _{\sigma _2}^q
\beta _{mn}\partial _{\sigma _1}^{m-p}\partial _{\sigma _2}^{n-q}\biggr),}
and the derivatives act on the right.
{}From Eq. \edg \ we obtain
\eqn\edh{{\cal P}_\mu ={\cal Q}^{-1}\bigl(\dot X_\mu -
\sum _{m,n\ge 0}\alpha_{mn}X_\mu ^{m,n}\bigr).}
Incerting the momentum from \edh \ into \eda \ we find
\eqn\edi{{\cal L}={\lambda \over 2}\biggl(\dot X^2-X^2_{,\sigma _1}-
X^2_{,\sigma _2}\biggr)-
\sum _{m,n\ge 0}\biggl(\widetilde \alpha _{mn}\dot X \dot X^{(m,n)}-
\widetilde \beta _{mn}\dot XX^{(m,n)}-
\widetilde \gamma _{mn}XX^{(mn)}\biggr),}
where the kinetic term is separated formally. New gauge fields
$\widetilde \alpha , \dots $ are introduced instead of the infinite
series of the Lagrange multiplyers $\alpha , \dots $ and their
derivatives and $\lambda $ is a gauge invariant field whit spin
1. The explicite form of these functions in terms of $\alpha ,
\dots $ can be found by power decomposition
of the operator ${\cal Q}^{-1}$.

In order to gauge the configuration space Lagrangian \edi \ we
suppose that the multiplyers $\widetilde \alpha , \dots $ are
functions of the evolution parameter $\tau $ and the
spatial world-sheet coordinates ${\sigma }$. From the invariance
of the action \edi \ with respect to the local
$W_{1+\infty }$ transformations we obtain the tranformation laws
for the gauge fields:
\eqn\edj{\eqalign{\delta \widetilde \alpha _{mn} & =
-\lambda f_{mn}-
\sum _{k,l\ge 0}\sum _{p,q=0}^{k,l}\pmatrix{k \cr p \cr }
\pmatrix {l \cr q \cr }\biggl((-)^{k+l}\partial _{\sigma _1}^p
\partial _{\sigma _2}^q\bigl(f_{kl}\widetilde \alpha
_{m-k+p,n-l+q}\bigr) \cr
& +\widetilde \alpha _{k,l}\partial _{\sigma _1}^p
\partial _{\sigma _2}^qf_{m-k+p,n-l+q}\biggr), \cr
\delta {\widetilde \beta }_{mn} & =
-\lambda \dot f_{mn}+
\sum _{k,l\ge 0}\sum _{p,q=0}^{k,l}\pmatrix{k \cr p \cr }
\pmatrix {l \cr q \cr }\biggl((-)^{k+l}\partial _{\sigma _1}^p
\partial _{\sigma _2}^q\bigl(f_{kl}{\widetilde \beta }
_{m-k+p,n-l+q} \cr
& +\dot f_{kl}{\widetilde \alpha }_{m-k+p,n-l+q}\bigr)+
{\widetilde \alpha }_{k,l}\partial _{\sigma _1}^p
\partial _{\sigma _2}^q\dot f_{m-k+p,n-l+q}+
{\widetilde \beta }\partial _{\sigma _1}^p\partial _{\sigma _2}^q
f_{m-k+p,n-l+q}\biggr), \cr
\delta {\widetilde \gamma }_{mn} & =
-\lambda\biggl(\bigl(\partial _{\sigma _1}^2+
\partial _{\sigma _2}^2\bigr)f_{mn}
+2\bigl(\partial _{\sigma _1}f_{k-1,l}+
\partial _{\sigma _2}f_{k,l-1}\bigr)+
f_{m-2,n}-f_{m,n-2}\biggr) \cr
& +\sum _{k,l\ge 0}\sum _{p,q=0}^{k,l}\pmatrix{k \cr p \cr }
\pmatrix {l \cr q \cr }\biggl((-)^{k+l}\partial _{\sigma _1}^p
\partial _{\sigma _2}^q\bigl(f_{kl}{\widetilde \gamma }
_{m-k+p,n-l+q} \cr
& +\dot f_{kl}{\widetilde \beta }_{m-k+p,n-l+q}\bigr)+
{\widetilde \gamma }_{k,l}\partial _{\sigma _1}^p
\partial _{\sigma _2}^qf_{m-k+p,n-l+q}\biggr),\cr }}

According to Eq. \edk \ all the quanties $\dot XX^{m,n}$
\ (if $D=\infty $) \ as well as the quantities
$(\dot X\dot X^{m,n})$ \ and \ $(XX^{m,n})$ \ for which \
$m+n=2k$ are independent.  So the formula \edk \ shows  that
there exists an symmetry of the action \edi \ with respect of the
transformations of Stukelberg type \rBPSS
\eqn\edm{\eqalign{\tilde \delta \tilde \alpha _{2M-m+1,m} &
=u_{2M-m+1,m}, \cr
\tilde \delta \tilde \alpha _{2M-m,m} & = -\sum _{K,l}
C^{K,l}_{M,m}\partial _{\sigma _1}^{2(K-M)-l+m}
\partial _{\sigma _2}^{l-m}u_{2K+1,l}, \cr
\tilde \delta \tilde \gamma _{2M-m+1,m} & =v_{2M-m+1,m}, \cr
\tilde \delta \tilde \gamma _{2M-m,m} & = -\sum _{K,l}
C^{K,l}_{M,m}\partial _{\sigma _1}^{2(K-M)-l+m}
\partial _{\sigma _2}^{l-m}v_{2K+1,l}, \cr }}
where \ $u_{m,n}$ \ and \ $v_{m,n}$ \ are arbitrary functions.
This invariance allows us to choose the folowing gauge fixing
\eqn\edn{\eqalign{\tilde \alpha _{2K-l+1,l} & =0, \cr
\tilde \gamma _{2K-l+1,l} & =0. \cr }}
In this gauge the even spin quantities $(\dot X\dot X^{2K-l+1,l})$
 \ and \ $(XX^{2K-l+1,l})$ are canceled in the action \edi .
Then the Lagrangian  became
\eqn\edo{\eqalign{{\cal L} & =
{\lambda \over 2}\biggl(\dot X^2-X^2_{,\sigma _1}-
X^2_{,\sigma _2}\biggr) \cr
& -\sum _{m,n\ge 0}\widetilde \beta _{mn}(\dot XX^{m,n})+
\sum _{2K\ge l}\biggl(\widetilde \alpha _{2K-l,l}(\dot X \dot X^{2K-l,l})-
\widetilde \gamma _{K-l,l}XX^{2K-l,l}\biggr), \cr }}
which in the case \ $D=\infty $ \ contains only independent
quantities. In any other case only finite number of quantities
survives.

At the end we note that, the Hamiltonian approach applyied here loses
the manifest Lorentz covariance and leads to Hamilton gauge. For
instance, here appear three infinite sequences of gauge fields
instead of one vector gauge field sequence that appears in the
manifestly Lorentz covariant approach. The generalization for the
case of arbitrary C-S $p$-branes is straightforward. In that case
we have $p+1$ infinite sequences of gauge fields i.e.  one
sequence of $p+1$ vector potentials.

\noindent {\bf Acnowledgements.} The author would like to thank
to Prof. Abdus Salam, the International Atomic Energy Agency and
UNESCO for hospitality in ICTP in Trieste and to Dr. L. Nikolova
for a critical reading of the manuscript.

\listrefs
\bye